# Optical Modulation by Conducting Interfaces

Farhad Karimi, Sina Khorasani
*School of Electrical Engineering, Sharif University of Technology, Tehran, Iran*

*Abstract*— We analyze the interaction of a propagating guided electromagnetic wave with a quantum well embedded in a dielectric slab waveguide. First, we design a quantum well based on InAlGaAs compounds with the transition energy of 0.8eV corresponding to a wavelength of 1.55µm. By exploiting the envelope function approximation, we derive the eigenstates of electrons and holes and the transition dipole moments, through solution of the Luttinger Hamiltonian. Next, we calculate the electrical susceptibility of a three-level quantum system (as a model for the two-dimensional electron gas trapped in the waveguide), by using phenomenological optical Bloch equations. We show that the two-dimensional electron gas behaves as a conducting interface, whose conductivity can be modified by controlling the populations of electrons and holes the energy levels. Finally, we design a slab waveguide in which a guided wave with the wavelength of 1.55µm experiences a strong coupling to the conducting interface. We calculate the propagation constant of the wave in the waveguide subject to the conducting interface, by exploiting the modified transfer matrix method, and establish it linear dependence on the interface conductivity. By presenting a method for controlling the populations of electrons and holes, we design a compact optical modulator with an overall length of around 60µm.

*Index Terms*— Conducting Interfaces, Optical Modulation, Quantum Optics, Ultrastrong Coupling.

## I. INTRODUCTION

Recent developments of quantum confined compound semiconductors and low-dimensional structures [1] have enabled new capabilities in integrated optoelectronics [2]. Many such integrated devices are fabricated using layered deposition of various semiconductors [3], and many applications including spontaneous light emission, stimulated light generation, detection of light, and optical modulation and switching are demonstrated [4,5].

Most of these applications are realized using physical phenomena such as electrooptic, thermooptic, or magnetooptic effects to allow control over the propagation and transmission of optical wavefronts. Modulation speeds in excess of 50GHz have so far been achieved in the basic Mach-Zhender Modulator (MZM) configuration. While MZM could utilize virtually any physical effect to allow optical modulation, normally the linear electrooptic effect in bulk semiconductors is used. High speed modulation is also possible using internal modulation schemes, for instance, by direct current modulation [6,7]. However, most wideband applications require external modulation of light.

A common material which is widely used in this area is $LiNbO_3$. But there is a drawback in using MZMs as external modulators is that they normally require very long propagation lengths of the order of a few millimeters. This requirement makes the MZMs highly inappropriate for integrated fabrication. There is also another problem with routine material platforms, which happen to be incompatible with the standard III/V compound semiconductor technology.

Hence, one should look for an alternative physical phenomenon other than the above, which would enable strong modulation, ease of fabrication, compatibility with III/V optoelectronics integration, linear response, tunability, and possibility of modern quantum optical applications including quantum information processing and cavity quantum electrodynamics.

More than a decade ago, we coined the term *Conducting Interface* to a family of optoelectronic devices, in which the electrooptic effect is maintained by a surface accumulation or depletion of electric charges [8-14]. While nowadays, graphene might be regarded as the ultimate conducting interface, our initial speculation [8] had been directed towards the inversion and depletion layers, surface states [11], and in particular, the quantum confined Two-Dimensional Electron Gas (2DEG) in III/V heterostructures. We developed a Modified Transfer Matrix Method (MTMM) to efficiently model the family of layered structures with conducting interfaces [9] and also deduce a coupled mode theory [13] of waveguides with conducting interfaces. More recently [14], the usefulness of conducting interfaces in modeling plasmonic modes of thin-film photonic crystals has been also shown. Notably, several graphene-based optical and THz modulators have also been recently reported [15-17] which are essentially the same as our basic structure [8]. Also, the Intel's photonic group had utilized the same idea of using depletion layers in fabrication of its high-efficiency Silicon optical modulator [18].

The present research develops a rigorous theoretical infrastructure to study the 2DEG as a conducting interface. We start from a fundamental design of the QW using envelope function approximation method and Luttinger Hamiltonian [19,20]. We have successfully used this technique to study the strongly coupled photon-exciton interaction in photonic nanocavities [21] and also design of GaN light emitting diodes [22]. We then formulate and derive the complex-valued and dispersive interface optical conductivity of the QW using quantum mechanical optical Bloch equations. Using the MTMM [9], we then proceed to an electromagnetic design of a semiconductor waveguide at the telecommunication wavelength of 1.55μm, which is covered by the QW structure. The phase-retardation and absorption per unit length of the structure may be calculated and shown to be a strong function of the population of electron and hole states. As a result, we have been able to design a relatively compact Mach-Zhender modulator with a design arm length of 61μm. The compactness and high DC mobility of III/V compounds would imply unprecedented modulation speeds, well in excess of few 100 GHz.

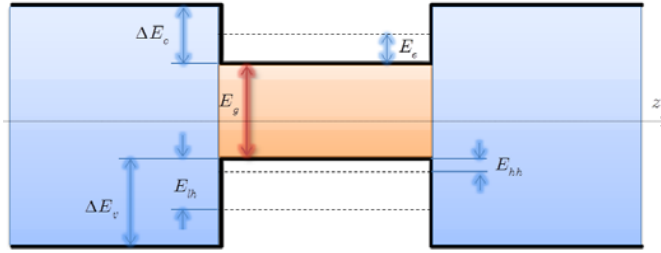

Fig. 1. A symmetric QW with the electron, heavy-hole and light-hole levels.

## II. Theory

### A. Quantum Well

Typical cross section of a symmetric QW is depicted in Fig. 1. Usually, QWs may have more than just the three confined levels as shown, however, a proper design will prohibit occurrence of unnecessary states. Temperature effects are here disregarded, and accuracy is maintained at low temperatures.

Conventional semiconductors such as InAs and GaAs with zinc-blend structure, have a non-degenerate conduction band in non-relativistic regime, and thus the electron states have S-like symmetries. On the other hand, the hole states in the valence bands are similar to P-orbitals, and not only the valence band is two-fold degenerate at the Γ-point (heavy-hole and light-hole bands), but also the split-off (SO) band may be close to these bands in some semiconductor band structures. Hereby, it can be deduced that for the electrons considering only one band in the envelope function approximation is sufficient, but for having enough precision in envelope function approximation considering at least two degenerate bands is required. More precision can be achieved by considering three bands as light-hole, heavy-hole and split-off.

S-like symmetry of electron states allows easy evaluation of the corresponding eigenvalues. For this purpose, the Schrödinger's equation with varying effective mass across the interfaces must be solved. The procedure is fairly standard and the governing characteristic equation may be easily found and numerically solved. But heavy-, light-hole, and SO states mix to form more complex orbitals than the simple P-like states $|X\uparrow\rangle, |X\downarrow\rangle, |Y\uparrow\rangle, |Y\downarrow\rangle, |Z\uparrow\rangle, |Z\downarrow\rangle$, and need to be dealt with differently.

But because of the symmetry in the *x-y* plane, and that also the calculations are done at the center of Brillouin Zone, so we can assume $k_\parallel^2 \cong 0$. This approximation is reasonably good at low temperatures. It can be easily deduced that the symmetry in the potential profile in heterostructure may result in $f_{\frac{3}{2},\frac{3}{2}} = f_{\frac{3}{2},-\frac{3}{2}}$ and $f_{\frac{3}{2},\frac{1}{2}} = f_{\frac{3}{2},\frac{-1}{2}}$, where $f_{i,j}$ denote the envelope functions. By defining $f_{hh} = f_{\frac{3}{2},\pm\frac{3}{2}}$ and $f_{lh} = f_{\frac{3}{2},\pm\frac{1}{2}}$, E, HH and LH states take the form

$$|\psi_e\rangle = \tfrac{1}{\sqrt{2}}|f_e\rangle(|S\uparrow\rangle + |S\downarrow\rangle)$$
$$|\psi_{hh}\rangle = \tfrac{1}{\sqrt{2}}|f_{hh}\rangle\left(\left|\tfrac{3}{2},\tfrac{3}{2}\right\rangle + \left|\tfrac{3}{2},\tfrac{-3}{2}\right\rangle\right) \quad (1)$$
$$|\psi_{lh}\rangle = \tfrac{1}{\sqrt{2}}|f_{lh}\rangle\left(\left|\tfrac{3}{2},\tfrac{1}{2}\right\rangle + \left|\tfrac{3}{2},\tfrac{-1}{2}\right\rangle\right)$$

Schrödinger equations describing the envelope functions are

$$\left\{\tfrac{-\hbar^2}{2}\nabla_z\cdot\left[\tfrac{1}{m_e}\nabla_z\right] + V_e(z)\right\}f_e(z) = E_n^e f_e(z)$$
$$\left\{\tfrac{-\hbar^2}{2m_0}\nabla_z\cdot\left[(\gamma_1 - 2\gamma_2)\nabla_z\right] + V_h(z)\right\}f_{hh}(z) = E_n^{hh} f_{hh}(z) \quad (2)$$
$$\left\{\tfrac{-\hbar^2}{2m_0}\nabla_z\cdot\left[(\gamma_1 + 2\gamma_2)\nabla_z\right] + V_h(z)\right\}f_{lh}(z) = E_n^{lh} f_{lh}(z)$$

By definition, $m_{hh}^* = \frac{1}{\gamma_1 - 2\gamma_2}m_0$ and $m_{lh}^* = \frac{1}{\gamma_1 + 2\gamma_2}m_0$ are HH and LH effective masses, respectively. These equations are also solved in a manner similar to that of electrons to find the envelope functions and energies.

Now, the dipole moments of basic transitions in the QW can be evaluated to find the Rabi frequencies. The HH-LH transition is forbidden by different symmetries of envelope functions, and therefore the non-zero Rabi frequencies are

$$G_h = \tfrac{1}{\hbar}\left(E_0 \hat{E}\right) \cdot \left\langle \psi_e \left| e\mathcal{R} \right| \psi_{hh} \right\rangle$$
$$G_l = \tfrac{1}{\hbar}\left(E_0 \hat{E}\right) \cdot \left\langle \psi_e \left| e\mathcal{R} \right| \psi_{lh} \right\rangle \quad (3)$$

Here, $G_h$ and $G_l$ are the Rabi frequencies of E-HH and E-LH transitions, respectively. Also, $\mathbf{E}=E_0 \hat{E}$ is the electric field and $\mathcal{D} = e\mathcal{R}$ is the dipole operator. By assuming that the variations of envelope functions are slow enough, we find

$$G_h = \tfrac{E_0}{2\hbar}\left\langle f_e | f_{hh}\right\rangle \left(\left\langle S\uparrow\right| + \left\langle S\downarrow\right|\right)\hat{E}\cdot e\mathcal{R}\left(\left|\tfrac{3}{2},\tfrac{3}{2}\right\rangle + \left|\tfrac{3}{2},\tfrac{-3}{2}\right\rangle\right)$$
$$G_l = \tfrac{E_0}{2\hbar}\left\langle f_e | f_{lh}\right\rangle \left(\left\langle S\uparrow\right| + \left\langle S\downarrow\right|\right)\hat{E}\cdot e\mathcal{R}\left(\left|\tfrac{3}{2},\tfrac{1}{2}\right\rangle + \left|\tfrac{3}{2},\tfrac{-1}{2}\right\rangle\right) \quad (4)$$

It should be considered that due to symmetry, only $\left\langle S|e\mathcal{X}|X\right\rangle$, $\left\langle S|e\mathcal{Y}|Y\right\rangle$ and $\left\langle S|e\mathcal{Z}|Z\right\rangle$ are non-zero, and all other matrix elements vanish. Also, only the transitions with the same spin are allowed. Besides, since the HH states have no image in $z$-direction. Hence, for TM mode in which $\hat{E}=\hat{z}$, $\left\langle \psi_e \left| \hat{E}\cdot e\mathcal{R} \right| \psi_{hh}\right\rangle$ identically vanishes. It means the E-HH transitions are not excited in TM mode, and the system reduces to a two-level one. Moreover, it has been shown previously that the TM polarization interacts much weaker with the conducting interface [9,10] and hence, we need to assume that the designed optoelectronic device should operate with a TE polarized excitation. Now, the image of dipole moments on the polarization of a TE mode EM field take the form

Table 1. Various selections for QW width $z_0$ and Al fraction $x$.

| $z_0$(nm) | $x$ | $E_g$ | $E_{hh}$ | $\Delta E_c$ | $\Delta E_v$ | $E_t$ |
|---|---|---|---|---|---|---|
| 2 | 0.09 | 46 | 15 | 48 | 18 | 799 |
| 3 | 0.1 | 47 | 14 | 54 | 20 | 799 |
| 5 | 0.14 | 50 | 11 | 75 | 28 | 799 |
| 6 | 0.18 | 52 | 9 | 97 | 36 | 799 |
| 7 | 0.26 | 52 | 9 | 140 | 52 | 799 |
| 8 | 0.42 | 53 | 8 | 226 | 84 | 799 |
| 9 | 0.90 | 54 | 7 | 486 | 180 | 799 |

$$\left\langle \psi_e \left| \hat{E}\cdot e\mathcal{R} \right| \psi_{hh}\right\rangle = \tfrac{1}{\sqrt{2}}\left\langle f_e | f_{hh}\right\rangle \left\langle S|e\mathcal{Y}|Y\right\rangle$$
$$\left\langle \psi_e \left| \hat{E}\cdot e\mathcal{R} \right| \psi_{lh}\right\rangle = \tfrac{1}{\sqrt{6}}\left\langle f_e | f_{lh}\right\rangle \left\langle S|e\mathcal{Y}|Y\right\rangle \quad (5)$$

Fortunately, the matrix elements in (5) are known [19,20]

$$\left\langle S|e\mathcal{Y}|Y\right\rangle = \frac{e\hbar}{\sqrt{2}E_t}\sqrt{\frac{E_G\left(E_G + \Delta_{SO}\right)}{m_e^*\left(E_G + \tfrac{2}{3}\Delta_{SO}\right)}} \quad (6)$$

Here, $E_t$ is transition energy being equal to $E_g + E_e + E_{hh}$ and $E_g + E_e + E_{lh}$, respectively for E-HH and E-LH transitions.

We design the QW in such a way that $E_t$ is equivalent to the energy of a photon with the communication wavelength 1.55μm. For reasons which will become clear later, we select this value smaller by 1meV, to avoid complete resonant excitation. We base our design on the quaternary system $In_{0.52}(Al_xGa_{1-x})_{0.48}As/In_{0.53}Ga_{0.47}As$. $In_{0.53}Ga_{0.47}As$ is a conventional material in fabrication of optical device, with a band gap energy of 738meV (corresponding to 1.68μm), referred to as standard InGaAs. This particular choice is favorable for our case, since for reaching the transition energy of 799 meV, sum of the first energies of E and HH levels should be 61meV. But if we use alloys such as InAs instead, the energy band gap is 360meV, then this value should be something around 440meV, demanding for either extremely thin QW width or large band offsets. In both cases, fabrication is much more difficult. Also, in the previous part, we saw that the structure with symmetry simplifies the calculation, so our design will be concurrent. Here, barriers are $In_{0.52}(Al_xGa_{1-x})_{0.48}As$, and the band offsets in such a heterostructure are $\Delta E_c=0.54x$(eV) and $\Delta E_v=0.20x$(eV) [23-26]. All remains is now to find the QW width $z_0$ in terms of the Al fraction $x$. Using the E, HH, and LH effective masses and band offset interpolation functions for $In_xAl_yGa_{1-x-y}As$ [27], we obtain the Table 1 as below. Here, we select $z_0$=9nm and $x$=0.9. Based on the lattice constant data, we find the lattice mismatch to be only 0.02%. The dipoles are now given as

$$D_{e-hh} = \left\langle \psi_e \left| \hat{E}\cdot e\mathcal{R} \right| \psi_{hh}\right\rangle = 26.15 \text{ Debye}$$
$$D_{e-lh} = \left\langle \psi_e \left| \hat{E}\cdot e\mathcal{R} \right| \psi_{lh}\right\rangle = 15.21 \text{ Debye} \quad (7)$$

### B. Susceptibility

The model Hamiltonian for the three-level Λ-system describing the QW is given simply by

$$\mathcal{H}_{sys} = E_e \left|e\right\rangle\left\langle e\right| + E_{g_{hh}}\left|g_{hh}\right\rangle\left\langle g_{hh}\right| + E_{g_{lh}}\left|g_{lh}\right\rangle\left\langle g_{lh}\right| \quad (8)$$

For simplicity, we denote the LH, HH, and E states as 1, 2, and 3, respectively. The interaction term is $\mathcal{H}_{int} = -\mathbf{E}(t)\cdot\mathcal{D}$, and hence we obtain

$$\mathcal{H}_{int} = -\hbar\cos(\omega t)\left(G_1|3\rangle\langle 1| + G_1^*|1\rangle\langle 3| + G_2|3\rangle\langle 2| + G_2^*|2\rangle\langle 3|\right) \quad (9)$$

with $G_1 \equiv \frac{1}{\hbar}E_0 D_{e-lh}$, $G_2 \equiv \frac{1}{\hbar}E_0 D_{e-hh}$, and $\omega$ being the angular frequency of optical wave. Total Hamiltonian is now

$$\mathcal{H}_{tot}(t) = \mathcal{H}_{sys} + \mathcal{H}_{int}(t) \quad (10)$$

If $\mathbf{D}(t) = \langle\mathcal{D}\rangle$ is the expected value of induced dipole in QW, then $\mathbf{P}(t) = N\mathbf{D}(t)$ is the macroscopic induced polarization, if $N$ is the density of carriers per unit volume. But due to definition, for the phasors of E-field and polarization we have

$$\mathcal{P} = \varepsilon_0 \chi(\omega)\mathcal{E} \quad (11)$$

Hence, we get the induced macroscopic polarization $\mathbf{P}(t)$

$$\mathbf{P}(t) = Re\{\mathcal{P}e^{j\omega t}\} = Re\{\varepsilon_0 \chi(\omega)\mathcal{E} e^{j\omega t}\} \quad (12)$$

Here, $\chi(\omega) = \chi_r(\omega) - j\chi_i(\omega)$, $\varepsilon(\omega) = \varepsilon_0\varepsilon_r(\omega)$ and $\varepsilon_r(\omega) = 1 + \chi(\omega)$ are susceptibility, electric permittivity and relative permittivity of medium, respectively. Therefore, the first step for calculating the susceptibility of the medium is to calculate the expected value of induced dipole in QW

$$\mathbf{D}(t) = \langle\mathcal{D}\rangle = \langle\psi|\mathcal{D}|\psi\rangle = Tr\{\hat{\rho}\mathcal{D}\} \quad (13)$$

$\hat{\rho} = |\psi\rangle\langle\psi|$ is the density matrix describing the system, and $|\psi\rangle$ is the ket of the system, which are found from

$$j\hbar\tfrac{\partial}{\partial t}|\psi\rangle = \mathcal{H}_{tot}|\psi\rangle, \quad j\hbar\tfrac{\partial}{\partial t}\hat{\rho} = [\mathcal{H}_{tot},\hat{\rho}] \quad (14)$$

After expanding in matrix form as $\hat{\rho} = [\rho_{ij}]$ and defining the transition frequencies $\hbar\omega_{ij} = E_i - E_j$, and $\xi_{13} = \rho_{11} - \rho_{33}$ and $\xi_{23} = \rho_{22} - \rho_{33}$ as population differences of electrons in different energy levels, we may obtain the time derivatives

$$\begin{aligned}
\tfrac{\partial}{\partial t}\xi_{23}(t) &= -j\cos(\omega t)\left(2G_2\rho_{23} - 2G_2^*\rho_{32} + G_1\rho_{13} - G_1^*\rho_{31}\right) \\
\tfrac{\partial}{\partial t}\xi_{13}(t) &= -j\cos(\omega t)\left(G_2\rho_{23} - G_2^*\rho_{32} + 2G_1\rho_{13} - 2G_1^*\rho_{31}\right) \\
\tfrac{\partial}{\partial t}\rho_{31}(t) &= -j\left[\omega_{31}\rho_{31} + \cos(\omega t)\left(G_1\rho_{33} - G_1\rho_{11} - G_2\rho_{21}\right)\right] \\
\tfrac{\partial}{\partial t}\rho_{32}(t) &= -j\left[\omega_{32}\rho_{32} + \cos(\omega t)\left(G_2\rho_{33} - G_2\rho_{22} - G_1\rho_{12}\right)\right] \\
\tfrac{\partial}{\partial t}\rho_{21}(t) &= -j\left[\omega_{21}\rho_{21} + \cos(\omega t)\left(G_2^*\rho_{31} - G_2\rho_{23}\right)\right]
\end{aligned} \quad (15)$$

Equations (15) explain time variations of the QW-field system. These are correct for an isolated system, but in reality isolation from the ambient is impossible. Coupling of dipole moments and lots of modes existing in free space causes inevitably spontaneous emission or absorption. Also, coherency is lost due to collisions and broadenings, so that by removing the EM field, off-diagonal elements of density matrix undergo exponential decay. These effects can be inserted into (15) phenomenologically. First, we associate time constants to each of these decays and/or losses as

$$\begin{aligned}
\tfrac{\partial}{\partial t}\xi_{23}(t) &= -\tfrac{1}{T_2}\left[\xi_{23}(t) - \bar{\xi}_{23}\right], \quad \tfrac{\partial}{\partial t}\xi_{13}(t) = -\tfrac{1}{T_1}\left[\xi_{13}(t) - \bar{\xi}_{13}\right] \\
\tfrac{\partial}{\partial t}\rho_{31}(t) &= -\tfrac{1}{T_{31}}\rho_{31} - j\omega_{31}\rho_{31}, \quad \tfrac{\partial}{\partial t}\rho_{32}(t) = -\tfrac{1}{T_{32}}\rho_{32} - j\omega_{32}\rho_{32} \\
\tfrac{\partial}{\partial t}\rho_{21}(t) &= -\tfrac{1}{T_{21}}\rho_{21} - j\omega_{21}\rho_{21}
\end{aligned} \quad (16)$$

where $\bar{\xi}_{i3}$ are mean values of excessive population of electrons in the absence of field, $T_i$ are the corresponding relaxation times, and also for off-diagonal elements of density matrix we have defined the $T_{ij}$ as the phase relaxation times. Now, it can be seen that the behavior of $\rho_{kl}$ in the absence of EM field is according to $\rho_{kl} = \bar{\rho}_{kl}\exp(-j\omega_{kl}t - T_{kl}^{-1})$. So, we may define the slow components as $\sigma_{kl} = \rho_{kl}e^{j\omega t}$ to obtain

$$\begin{aligned}
\tfrac{\partial}{\partial t}\xi_{23}(t) &= -\frac{\xi_{23}(t) - \bar{\xi}_{23}}{T_2} + \cos(\omega t)\left[2\operatorname{Im}\{G_2\sigma_{32}^* e^{j\omega t}\} + \operatorname{Im}\{G_1\sigma_{31}^* e^{j\omega t}\}\right] \\
\tfrac{\partial}{\partial t}\xi_{13}(t) &= -\frac{\xi_{13}(t) - \bar{\xi}_{13}}{T_1} + \cos(\omega t)\left[\operatorname{Im}\{G_2\sigma_{32}^* e^{j\omega t}\} + 2\operatorname{Im}\{G_1\sigma_{31}^* e^{j\omega t}\}\right] \\
\tfrac{\partial}{\partial t}\sigma_{31}(t) &= -\frac{\sigma_{31}}{T_{31}} + j(\omega - \omega_{31})\sigma_{31} - j\cos(\omega t)e^{j\omega t}\left(G_1\rho_{33} - G_1\rho_{11} - G_2\sigma_{21}e^{-j\omega t}\right) \\
\tfrac{\partial}{\partial t}\sigma_{32}(t) &= -\frac{\sigma_{32}}{T_{32}} + j(\omega - \omega_{32})\sigma_{32} - j\cos(\omega t)e^{j\omega t}\left(G_2\rho_{33} - G_2\rho_{22} - G_1\sigma_{12}e^{j\omega t}\right) \\
\tfrac{\partial}{\partial t}\sigma_{21}(t) &= -\frac{\sigma_{21}}{T_{21}} + j(\omega - \omega_{21})\sigma_{21} - j\cos(\omega t)e^{j\omega t}\left(G_2^*\sigma_{31}e^{-j\omega t} - G_2\sigma_{23}e^{j\omega t}\right)
\end{aligned} \quad (17)$$

For a Λ-configuration we have $\sigma_{12} \cong 0$, and also rapidly oscillating terms on the right-hand-side of (17) with frequencies $\pm 2\omega$ may be dropped. The steady-state solution can be then found by setting $\frac{\partial}{\partial t} \equiv 0$. After some algebra and defining $\Omega_{3i} = T_{3i}^{-1} |G_i^2| [T_{3i}^{-2} + (\omega - \omega_{3i})^2]^{-1}$ we can find

$$\tilde{\xi}_{13} = \frac{(1 + T_2\Omega_{32})\bar{\xi}_{13} - \frac{1}{2}T_1\Omega_{32}\bar{\xi}_{23}}{(1 + T_1\Omega_{31})(1 + T_2\Omega_{32}) - \frac{1}{4}T_1T_2\Omega_{32}\Omega_{31}}$$

$$\tilde{\xi}_{23} = \frac{(1 + T_1\Omega_{31})\bar{\xi}_{23} - \frac{1}{2}T_2\Omega_{31}\bar{\xi}_{13}}{(1 + T_1\Omega_{31})(1 + T_2\Omega_{32}) - \frac{1}{4}T_1T_2\Omega_{32}\Omega_{31}} \quad (18)$$

$$\tilde{\sigma}_{31} = \frac{j\frac{1}{2}G_1}{T_{31}^{-1} - j(\omega - \omega_{31})} \frac{(1 + T_2\Omega_{32})\bar{\xi}_{13} - \frac{1}{2}T_1\Omega_{32}\bar{\xi}_{23}}{(1 + T_1\Omega_{31})(1 + T_2\Omega_{32}) - \frac{1}{4}T_1T_2\Omega_{32}\Omega_{31}}$$

$$\tilde{\sigma}_{32} = \frac{j\frac{1}{2}G_2}{T_{32}^{-1} - j(\omega - \omega_{32})} \frac{(1 + T_1\Omega_{31})\bar{\xi}_{23} - \frac{1}{2}T_2\Omega_{31}\bar{\xi}_{13}}{(1 + T_1\Omega_{31})(1 + T_2\Omega_{32}) - \frac{1}{4}T_1T_2\Omega_{32}\Omega_{31}}$$

Defining $N_s$ as carrier surface density, we get the polarization

$$\mathbf{P}(t) = \mathrm{Re}\{\varepsilon_0 \chi(\omega) \mathbf{E} e^{j\omega t}\} = N_s \mathbf{D}(t) = N_s Tr\{\hat{\rho}\mathcal{D}\}$$
$$= 2N_s \mathrm{Re}\{(\tilde{\sigma}_{31}\mathbf{D}_{31} + \tilde{\sigma}_{32}\mathbf{D}_{32})^* e^{j\omega t}\} \quad (19)$$

From this result we may obtain the following

$$\chi^*(\omega) = \frac{2N_s}{\varepsilon_0 \hbar} \left[ \frac{j(\mathbf{E} \cdot \mathbf{D}_{31})^2}{T_{31}^{-1} - j(\omega - \omega_{31})} \tilde{\xi}_{13} + \frac{j(\mathbf{E} \cdot \mathbf{D}_{32})^2}{T_{32}^{-1} - j(\omega - \omega_{32})} \tilde{\xi}_{23} \right] \quad (20)$$

Expanding (20) as $\chi(\omega) = \chi^{(1)}(\omega) + \chi^{(3)}(\omega)|\mathbf{E}|^2$ and after considerable algebra we finally arrive at

$$\chi_r^{(1)}(\omega) = -\frac{N_s}{\hbar\varepsilon_0} \left[ \frac{(\hat{E} \cdot \mathbf{D}_{31})^2 (\omega - \omega_{31})}{T_{31}^{-2} + (\omega - \omega_{31})^2} \bar{\xi}_{13} + \frac{(\hat{E} \cdot \mathbf{D}_{32})^2 (\omega - \omega_{32})}{T_{32}^{-2} + (\omega - \omega_{32})^2} \bar{\xi}_{23} \right]$$

$$\chi_i^{(1)}(\omega) = \frac{N_s}{\hbar\varepsilon_0} \left[ \frac{T_{31}^{-1}(\hat{E} \cdot \mathbf{D}_{31})^2}{T_{31}^{-1} + (\omega - \omega_{31})^2} \bar{\xi}_{13} + \frac{T_{32}^{-1}(\hat{E} \cdot \mathbf{D}_{32})^2}{T_{32}^{-1} + (\omega - \omega_{32})^2} \bar{\xi}_{23} \right] \quad (21)$$

Expressions for $\chi^{(3)}(\omega)$ are similarly found and given in the Appendix A. Figure 2 illustrates the behavior of $\chi^{(1)}(\omega)$ under thermal equilibrium. It is assumed that $T_{31}=T_{32}=T_{\mathrm{phase}}$, where the phase relaxation time $T_{\mathrm{phase}}$ of a free charge carrier is quite short (about 100fs) compared to that of E-H pairs in the bulk of QWs in low temperatures, which is about few picoseconds. Time-domain coherent optical experiments show a big range for $T_{\mathrm{phase}}$ [28-33]. Strongly interacting E-H pairs excited to high densities by band-to-band transitions show $T_{\mathrm{phase}}$ of the order of a few 100fs [24]. Photon echo experiments on impurity-related bound excitons have revealed a much weaker exciton-exciton interaction with corresponding relaxation times up to 100ps [28,33,34]. Here, we assume $T_{\mathrm{phase}}$=100ps.

The individual population of the three E, HH, and LH states are functions of bias current as well as a possible external optical pumping. This has been modeled in Appendix B. The complete rate equation model could be solved only if all parameters are known, and for the present study we limit ourselves to the case of population inversion exactly at which the absorption is minimal. For different bias currents and/or pumping levels, optical absorption or gain should be in principle achievable. These are well beyond the scope of this paper and will be the subject of a future study.

*C. Interface Conductivity*

Based on Ohm's law, for the phasors of surface current density $\mathcal{J}$ and electric field $\mathcal{E}$ we must have

$$\mathcal{J} = \sigma_s \mathcal{E} \quad (22)$$

But the surface current density $\mathcal{J}$ consists of surface current density due to freely moving charges $\mathcal{J}_f$ and surface current density caused by bound charges $\mathcal{J}_p$. We normally have no free charges moving in parallel to the interface, while the contribution of the trapped charges in QW may be regarded only as the $\mathcal{J}_p$ caused by E-H pairs, given by

$$\mathcal{J}_p = \frac{\partial}{\partial t}\mathcal{P} \quad (23)$$

with $\mathcal{P}$ being the macroscopic polarization phasor, and is equal to $\varepsilon_0 \chi(\omega) \mathcal{E}$. Hence, we find the expression for the interface conductivity of the 2DEG as

$$\sigma_s = j\omega\varepsilon_0 \chi^{(1)}(\omega) \quad (24)$$

which may be directly evaluated by plugging (21) in the right-hand-side of (24). Since $\sigma_s$ is given in the units of Siemens, it is appropriate to normalize it as $\eta_0 \sigma_s$, where $\eta_0$ is the intrinsic impedance of free space.

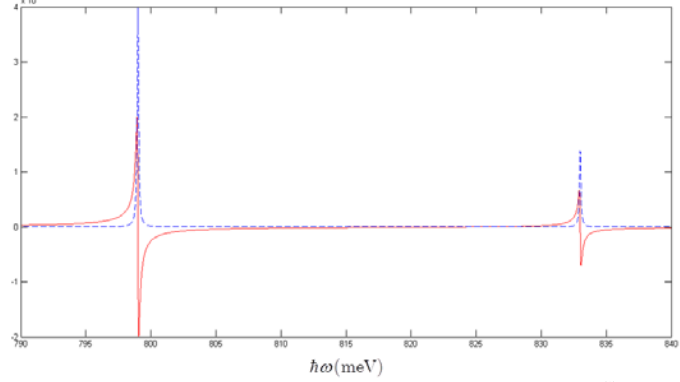

Fig. 2. Real part (Solid line) and imaginary part (Dashed line) of $\chi^{(1)}(\omega)$.

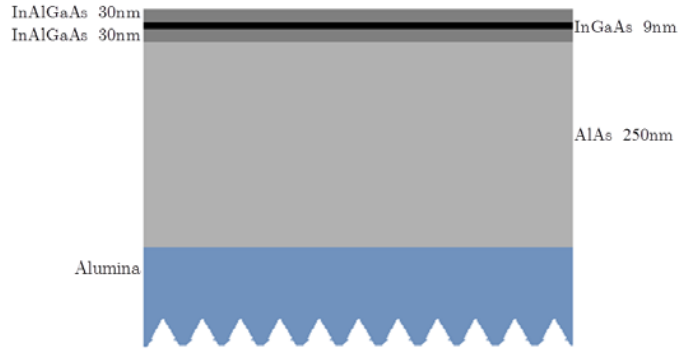

Fig. 3. The cross section of optical waveguide. Propagation direction is normal to the paper.

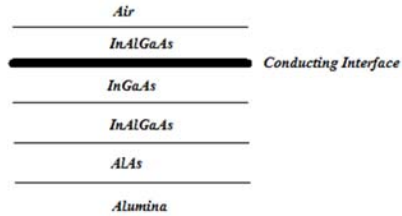

Fig. 4. The optical layered structure used in the analysis of proposed waveguide.

### D. Waveguide Design

So far we have been engaged in the design of QW and derivation of the corresponding interface conductivity in terms of various parameters. In practice, the surface concentration of carriers in the 2DEG can be varied over several orders of magnitude within $10^{11}$-$10^{13}$cm$^{-2}$ simply by applying a gate control voltage similar to the High-Electron Mobility Transistors (HEMTs). This allows a strong external control on the interface conductivity and therefore the associate electrooptic effect. Furthermore, the presence of a second light beam in resonance with the other transition, namely being the E-LH, gives rise to another strong control mechanism, enabling all-optical switching of light. It is not the purpose of the present study to discuss the latter idea, and we only limit ourselves to the first, that is the electrooptic modulation through interface density control of 2DEG. Such phase modulators could be as fast as comparable HEMT technologies. Therefore, modulation/switching bandwidths in excess of a few 100GHz should be well in reach for sufficiently compact and optimized designs.

## III. RESULTS

Our proposed optical modulator is actually a simple semiconductor optical waveguide, capped with the QW structure designed earlier to confine a 2DEG at the top, functioning as the conducting interface. The electromagnetic design of such layered structure could be done by combining the MTMM [9] with either a root search algorithm, or the our previously reported variational approach [35] for extraction of eigemodes. Details of calculations are fairly standard and dropped here for the sake of brevity.

The typical structure is illustrated in Fig. 3. The Alumina substrate provides the low-index material as optical substrate and also the insulating mechanical basement. The 250nm-thick AlAs is both the main dielectric of the waveguide in which the light is essentially confined, and a material to relax the lattice mismatch between the substrate and the QW. Here, the $In_{0.53}Ga_{0.47}As$ layer with the width of 9nm has the biggest refractive index of 3.73. The barrier layers are symmetric and chosen to be $In_{0.52}Al_{0.432}Ga_{0.048}As$ with the width of 30nm and refractive index of 3.585. The refractive indices of AlAs and Alumina are 2.9 and 1.77, respectively. The cover of the waveguide is simply assumed to be air. We model the proposed waveguide as the layered structure in Fig. 3, with one conducting interface accounting for the optical effect of the 2DEG. We also have studied the effect of non-zero thickness of the conducting interface, which was noticed to be quite negligible for the TE excitation. Hence, the zero-thickness conducting interface is a very efficient and convenient representation of the actual 2DEG's optical contribution. We calculated the TE-mode profile of the first guided mode which is plotted in Fig. 5.

By using the method of analysis as described above, the propagation constant can be calculated for the waveguide demonstrated in Fig. 3, for different values of interface conductivity. We refer to the propagation constant in the case of absence of conducting interface as $\beta_0$, which is found to be $8.03 \times 10^6 m^{-1}$. In Fig. 5, the intensity distribution of the electric field is plotted. The conducting interface is located at $x=0$. If we consider that the conducting interface with conductivity $\sigma$ would correspond to the propagation constant $\beta_\sigma$, then we may define the phase modulation per unit propagation of length as $\Delta(\Delta\varphi / \Delta z) \equiv \frac{360}{2\pi}(\beta_\sigma - \beta_0)\frac{°}{\mu m}$, and is plotted for different values of $\sigma$ as shown in Fig. 6.

In Fig. 6, the phase modulation per unit length $\Delta(\Delta\varphi / \Delta z)$ is plotted for different values of surface carrier density $N_s$ of $0.3 \times 10^{12} cm^{-2}$, $10^{12} cm^{-2}$, and $3 \times 10^{12} cm^{-2}$. Furthermore, $T_{phase}$ is taken to be 100ps with $\bar{\xi}_{13} = 0.5$. Also, $\bar{\xi}_{23}$ varies between 0.05 and 0.49. These particular choice of steady state populations lead to vanishing real part of the interface conductivity at the wavelength of interest, corresponding to a change in the interface conductivity from $-j0.0097\eta_0^{-1}$ to $-j0.105\eta_0^{-1}$.

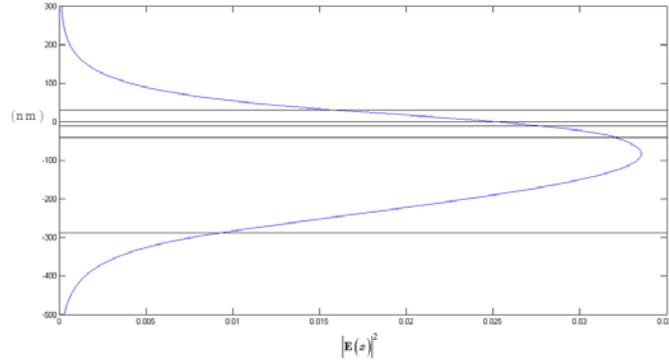

Fig. 5. Calculated intensity profile of the zeroth-order guided TE-mode for the structure defined in Figs. 3 and 4.

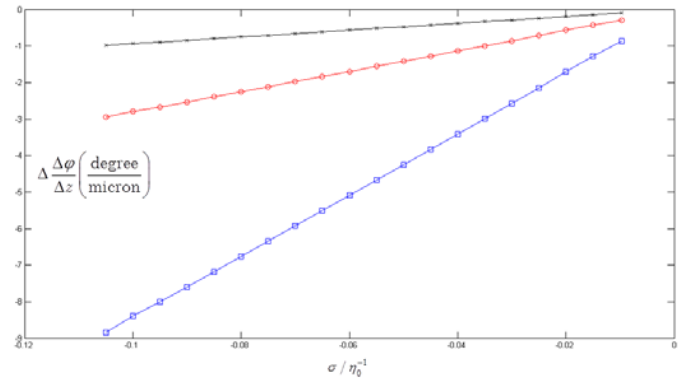

Fig. 6. Illustration of linear relationship between the phase modulation per unit propagation length versus normalized interface conductivity for various surface carrier concentrations of $3 \times 10^{11} cm^{-2}$ (x), $10^{12} cm^{-2}$ (○), and $3 \times 10^{12} cm^{-2}$ (□) (the real part of $\sigma$ is effectively zero).

By inspecting Fig. 6 it can be deduced that the phase modulation per unit length $\Delta(\Delta\varphi / \Delta z)$ is linearly dependent on the conductivity of conducting interface. As a result, in the range of interest, $\Delta(\Delta\varphi / \Delta z)$ varies between $-0.29°/\mu m$ to $-2.95°/\mu m$. If we set the target total phase modulation to be equal to $-180°$, then the needed length of the propagation should be only 61μm. Therefore, if we have two waveguide arms with the same length of 61μm in a Mach-Zhender configuration, one with conducting interface and the other one without conducting interface, then by propagating two wave with equal initial phases in each arm, the final phase of waves may differ up to 180°. The interface conductivity can be easily controlled by applying a bias voltage to control the population of electrons at different energy levels.

## IV. CONCLUSION

We studied the possibility of ultrastrong coupling in an optical waveguide operating in optical wavelengths. For this purpose, first we designed a quantum well with the transition optical wavelength of 1.55μm. We calculated the energy eigenstates and transition dipole moments. The quantum well was modeled with a Λ-shaped three-level quantum system, and its susceptibility was found by solving optical Bloch equations. We showed that due to the interaction of electromagnetic field with the well, a conducting interface appears, whose conductivity and its dependence on the population of different states was calculated. Finally, we designed a waveguide enclosing the well, in which a strong coupling between the guided wave and conducting interface was expected. By exploiting modified transfer matrix method, the propagation constant in the designed waveguide with conducting interface was calculated. It was observed that propagation constant had a linear dependence on the conductivity. By offering a method for controlling the population at different energy levels of quantum well, we designed a Mach-Zhender optical modulator with a length of 61μm, that was significantly smaller in comparison with most other external optical modulators.

## APPENDIX A: 3$^{RD}$-ORDER SUSCEPTIBILITY

Following the same method explained under (26) we may obtain the real and imaginary parts as:

$$\chi_r^{(3)}(\omega) = \frac{N_s}{\hbar^3 \varepsilon_0} \times$$

$$\left\{ \frac{(\hat{E} \cdot \mathbf{D}_{31})^2 (\omega - \omega_{31})}{T_{31}^{-2} + (\omega - \omega_{31})^2} \left[ \frac{1}{2} \frac{T_1 T_{32}^{-1} (\hat{E} \cdot \mathbf{D}_{32})^2}{T_{32}^{-2} + (\omega - \omega_{32})^2} \bar{\xi}_{23} + \frac{T_1 T_{31}^{-1} (\hat{E} \cdot \mathbf{D}_{31})^2}{T_{31}^{-2} + (\omega - \omega_{31})^2} \bar{\xi}_{13} \right] \right.$$

$$\left. + \frac{(\hat{E} \cdot \mathbf{D}_{32})^2 (\omega - \omega_{32})}{T_{32}^{-2} + (\omega - \omega_{32})^2} \left[ \frac{T_2 T_{32}^{-1} (\hat{E} \cdot \mathbf{D}_{32})^2}{T_{32}^{-2} + (\omega - \omega_{32})^2} \bar{\xi}_{23} + \frac{1}{2} \frac{T_2 T_{31}^{-1} (\hat{E} \cdot \mathbf{D}_{31})^2}{T_{31}^{-2} + (\omega - \omega_{31})^2} \bar{\xi}_{13} \right] \right\}$$

$$\chi_i^{(3)}(\omega) = \frac{-N_s}{\hbar^3 \varepsilon_0} \times$$

$$\left\{ \frac{T_{31}^{-1} (\hat{E} \cdot \mathbf{D}_{31})^2}{T_{31}^{-2} + (\omega - \omega_{31})^2} \left[ \frac{1}{2} \frac{T_1 T_{32}^{-1} (\hat{E} \cdot \mathbf{D}_{32})^2}{T_{32}^{-2} + (\omega - \omega_{32})^2} \bar{\xi}_{23} + \frac{T_1 T_{31}^{-1} (\hat{E} \cdot \mathbf{D}_{31})^2}{T_{31}^{-2} + (\omega - \omega_{31})^2} \bar{\xi}_{13} \right] \right.$$

$$\left. + \frac{T_{32}^{-1} (\hat{E} \cdot \mathbf{D}_{32})^2}{T_{32}^{-2} + (\omega - \omega_{32})^2} \left[ \frac{T_2 T_{32}^{-1} (\hat{E} \cdot \mathbf{D}_{32})^2}{T_{32}^{-2} + (\omega - \omega_{32})^2} \bar{\xi}_{23} + \frac{1}{2} \frac{T_2 T_{31}^{-1} (\hat{E} \cdot \mathbf{D}_{31})^2}{T_{31}^{-2} + (\omega - \omega_{31})^2} \bar{\xi}_{13} \right] \right\}$$

Typical values for $T_{1,2}$ are in the range of 1-10ns.

## APPENDIX B: SIMPLE RATE EQUATION MODEL FOR THREE-LEVEL SURFACE SYSTEM

We assume that three E, HH, and LH energy levels form a Λ-transition structure, with populations affected by bias current, non-radiative recombinations (carrier decay), and radiative recombinations. Radiative recombinations may be influenced by the presence of interacting photons, the density of which may be expressed in a fourth equation.

### B.1. Modulation

If the radiation intensity is too small to trigger the stimulated emission, and if we also disregard the spontaneous radiative recombinations, then the photon rate equation decouples completely from the carrier rate equations. Now, we now may write down the phenomenological equations

$$\frac{dN_e}{dt} = \eta_e \frac{I}{qA} - \frac{|d_{eh}|^2 N_e N_h}{q^2 \tau_{eh}} - \frac{|d_{el}|^2 N_e N_l}{q^2 \tau_{el}}$$

$$\frac{dN_h}{dt} = \eta_h \frac{I}{qA} - \frac{|d_{eh}|^2 N_e N_h}{q^2 \tau_{eh}}, \quad \frac{dN_l}{dt} = \eta_l \frac{I}{qA} - \frac{|d_{el}|^2 N_e N_l}{q^2 \tau_{el}}$$

Here, $q$ is the electronic charge (positive sign taken for both charges), and $\eta_e$, $\eta_h$, and $\eta_l$ are respectively is E, HH, and LH injection efficiencies. $A$ is the top gated area of well layer in which recombinations take place. Also, $\tau_{eh}$ and $\tau_{el}$ are respective radiative lifetimes of E-HH and E-LH band-to-band transitions, while $d_{eh}$ and $d_{el}$ are the associated quantum mechanical dipoles.

Here, we have disregarded all other non-radiative carrier loss mechanism (which may be safely dropped for high quality grown samples). These rate equations are evidently nonlinear. We now define the non-dimensional linear differential gains

$$g_{eh}(N_e) = \frac{|d_{eh}|^2}{q^2} N_e, \quad g_{el}(N_e) = \frac{|d_{el}|^2}{q^2} N_e$$

to obtain

$$\frac{dN_e}{dt} = \eta_e \frac{I}{qA} - g_{eh}(N_e) \frac{N_h}{\tau_{eh}} - g_{el}(N_e) \frac{N_l}{\tau_{el}}$$

$$\frac{dN_h}{dt} = \eta_h \frac{I}{qA} - g_{eh}(N_e) \frac{N_h}{\tau_{eh}}, \quad \frac{dN_l}{dt} = \eta_l \frac{I}{qA} - g_{el}(N_e) \frac{N_l}{\tau_{el}}$$

Charge neutrality condition requires that

$$\frac{dN_e}{dt} = \frac{dN_h}{dt} + \frac{dN_l}{dt}$$

This makes one of the three equations dependent on the two other. Therefore, we have two nonlinear equations in terms of three unknowns. Hence, we arrive at the condition $\eta_e = \eta_h + \eta_l < 1$. It should be noted that for high-quality samples, the electron injection efficiency $\eta_e$ and therefore the sum $\eta_h + \eta_l$ are sufficiently close to unity. Under steady state bias with $d/dt = 0$ we obtain the under-determined system

$$\eta_e \frac{I}{qA} = g_{eh}(N_e) \frac{N_h}{\tau_{eh}} + g_{el}(N_e) \frac{N_l}{\tau_{el}}$$

$$\eta_h \frac{I}{qA} = g_{eh}(N_e) \frac{N_h}{\tau_{eh}}, \quad \eta_l \frac{I}{qA} = g_{el}(N_e) \frac{N_l}{\tau_{el}}$$

The solution is indeterminate unless we incorporate at least one non-radiative carrier loss mechanism. Then, we obtain

$$\frac{dN_e}{dt} = \eta_e \frac{I}{qA} - g_{eh}(N_e) \frac{N_h}{\tau_{eh}} - g_{el}(N_e) \frac{N_l}{\tau_{el}} - \frac{N_e}{\tau_e}$$

$$\frac{dN_h}{dt} = \eta_h \frac{I}{qA} - g_{eh}(N_e) \frac{N_h}{\tau_{eh}} - \frac{N_h}{\tau_h}$$

$$\frac{dN_l}{dt} = \eta_l \frac{I}{qA} - g_{el}(N_e) \frac{N_l}{\tau_{el}} - \frac{N_l}{\tau_l}$$

in which $\tau_e$, $\tau_h$, and $\tau_l$ are respectively for E, HH, and LH non-radiative lifetimes. Charge neutrality now reads

$$(\eta_e - \eta_h - \eta_l) \frac{I}{qV} = + \frac{N_e}{\tau_e} - \frac{N_h}{\tau_h} - \frac{N_l}{\tau_l}$$

If the left hand side is to be zero, then we may still take advantage of the equation $\eta_e = \eta_h + \eta_l < 1$ to arrive at

$$\frac{N_e}{\tau_e} = \frac{N_h}{\tau_h} + \frac{N_l}{\tau_l}$$

Under steady-state the rate equations transform into another under-determined system with no solutions. Otherwise, it might be well that $\eta_e \neq \eta_h + \eta_l$ and steady state solutions to the triple rate equation should be sought numerically.

$$\eta_e \frac{I}{qA} = g_{eh}(N_e) \frac{N_h}{\tau_{eh}} + g_{el}(N_e) \frac{N_l}{\tau_{el}} + \frac{N_e}{\tau_e}$$

$$\eta_h \frac{I}{qA} = g_{eh}(N_e) \frac{N_h}{\tau_{eh}} + \frac{N_h}{\tau_h}, \quad \eta_l \frac{I}{qA} = g_{el}(N_e) \frac{N_l}{\tau_{el}} + \frac{N_l}{\tau_l}$$

These equations offer solutions in terms of the solutions of three third-order algebraic equations. Expressions are not useful, and numerical solutions are much better suited to this problem. If we neglect the dependences of non-dimensional gains on their arguments, we arrive at the solutions

$$N_e = \frac{\tau_e}{qA}\left[\eta_e - \eta_h\left(1+\frac{\tau_{eh}}{g_{eh}\tau_h}\right)^{-1} - \eta_l\left(1+\frac{\tau_{el}}{g_{el}\tau_l}\right)^{-1}\right]I$$

$$N_h = \frac{\eta_h}{qA}\left(\frac{g_{eh}}{\tau_{eh}}+\frac{1}{\tau_h}\right)^{-1}I, \quad N_l = \frac{\eta_l}{qA}\left(\frac{g_{el}}{\tau_{el}}+\frac{1}{\tau_l}\right)^{-1}I$$

*B.2. Light Generation*

If the structure is to be used for generation of light (spontaneous emission, or lasing), then a fourth equation describing photons is needed.

*B.2.1. Spontaneous Emission*

For the case of spontaneous emission we have

$$\frac{dN_e}{dt} = \eta_e \frac{I}{qA} - g_{eh}(N_e)\frac{N_h}{\tau_{eh}} - g_{el}(N_e)\frac{N_l}{\tau_{el}} - \frac{N_e}{\tau_e}$$

$$\frac{dN_h}{dt} = \eta_h \frac{I}{qA} - g_{eh}(N_e)\frac{N_h}{\tau_{eh}} - \frac{N_h}{\tau_h}$$

$$\frac{dN_l}{dt} = \eta_l \frac{I}{qA} - g_{el}(N_e)\frac{N_l}{\tau_{el}} - \frac{N_l}{\tau_l}$$

$$\frac{dS}{dt} = g_{eh}(N_e)\frac{N_h}{\tau_{eh}} - \frac{S}{\tau_p}$$

With the steady-state solutions

$$N_e = \frac{\tau_e}{qA}\left[\eta_e - \eta_h\left(1+\frac{\tau_{eh}}{g_{eh}\tau_h}\right)^{-1} - \eta_l\left(1+\frac{\tau_{el}}{g_{el}\tau_l}\right)^{-1}\right]I$$

$$N_h = \frac{\eta_h}{qA}\left(\frac{g_{eh}}{\tau_{eh}}+\frac{1}{\tau_h}\right)^{-1}I, \quad N_l = \frac{\eta_l}{qA}\left(\frac{g_{el}}{\tau_{el}}+\frac{1}{\tau_l}\right)^{-1}I$$

$$S = g_{eh}(N_e)\frac{\eta_h \tau_p}{qA\tau_{eh}}\left(\frac{g_{eh}}{\tau_{eh}}+\frac{1}{\tau_h}\right)^{-1}I$$

*B.2.2. Lasing*

For the case of lasing, rate equations must be reorganized as

$$\frac{dN_e}{dt} = \eta_e \frac{I}{qA} - v_g g(N_e, N_h)S - g_{el}(N_e)\frac{N_l}{\tau_{el}} - \frac{N_e}{\tau_e}$$

$$\frac{dN_h}{dt} = \eta_h \frac{I}{qA} - v_g g(N_e, N_h)S - \frac{N_h}{\tau_h}$$

$$\frac{dN_l}{dt} = \eta_l \frac{I}{qA} - g_{el}(N_e)\frac{N_l}{\tau_{el}} - \frac{N_l}{\tau_l}$$

$$\frac{dS}{dt} = \Gamma v_g g(N_e, N_h)S - \frac{S}{\tau_p}$$

Where we have taken note of the fact that photons having the surface density $S$, correspond only to the E-HH transitions. Here, $v_g$ and $\Gamma$ are respectively the group velocity of photons, and the non-dimensional confinement factor and also $g(N_e, N_h)$ is some appropriate optical gain expression along the propagation length, having the unit of inverse length. In this system, the spontaneous emission term is dropped. If we take care of this last remaining term, we arrive at

$$\frac{dN_e}{dt} = \eta_e \frac{I}{qA} - v_g g(N_e, N_h) S - g_{eh}(N_e) \frac{N_h}{\tau_{eh}} - g_{el}(N_e) \frac{N_l}{\tau_{el}} - \frac{N_e}{\tau_e}$$

$$\frac{dN_h}{dt} = \eta_h \frac{I}{qA} - v_g g(N_e, N_h) S - g_{eh}(N_e) \frac{N_h}{\tau_{eh}} - \frac{N_h}{\tau_h}$$

$$\frac{dN_l}{dt} = \eta_l \frac{I}{qA} - g_{el}(N_e) \frac{N_l}{\tau_{el}} - \frac{N_l}{\tau_l}$$

$$\frac{dS}{dt} = \Gamma v_g g(N_e, N_h) S + g_{eh}(N_e) \frac{N_h}{\tau_{eh}} - \frac{S}{\tau_p}$$

If we disregard the dependences of non-dimensional and optical gain expressions on their arguments, we arrive at the steady-state solution at high intensities from the last equation

$$g(N_e, N_h) \approx (\Gamma v_g \tau_p)^{-1}$$

From which we get

$$\eta_e \frac{I}{qA} = \frac{1}{\Gamma \tau_p} S + g_{eh} \frac{N_h}{\tau_{eh}} + g_{el} \frac{N_l}{\tau_{el}} + \frac{N_e}{\tau_e}$$

$$\eta_h \frac{I}{qA} = \frac{1}{\Gamma \tau_p} S + g_{eh} \frac{N_h}{\tau_{eh}} + \frac{N_h}{\tau_h}, \quad \eta_l \frac{I}{qA} = g_{el} \frac{N_l}{\tau_{el}} + \frac{N_l}{\tau_l}$$

$$\frac{S}{\tau_p} = \Gamma v_g g(N_e, N_h) S + g_{eh}(N_e) \frac{N_h}{\tau_{eh}}$$

With neglection of spontaneous emission, these would offer the solutions

$$N_e = \tau_e \left[ \eta_e \frac{1}{qA} I - \frac{S}{\tau_p} \right] - \frac{\tau_e g_{el} \eta_l}{\tau_{el} qA} \left( \frac{1}{\tau_e} + \frac{1}{\tau_{el}} \right)^{-1} I$$

$$N_l = \eta_l \frac{1}{qA} \left( \frac{1}{\tau_e} + \frac{1}{\tau_{el}} \right)^{-1} I, \quad N_h = \tau_h \left( \eta_h \frac{1}{qA} I - \frac{S}{\tau_p} \right)$$

Clearly, there is a threshold current $I_{th}$ at which spontaneous and stimulated emissions are equal. Hence, we have

$$\Gamma v_g g(N_e, N_h) S_{th} = g_{eh}(N_e) \frac{N_h}{\tau_{eh}}$$

where $S_{th}$ is the laser photon density at the onset of stimulated emission. On the other hand, we must have

$$\eta_h \frac{1}{qA} I_{th} \geq \frac{S_{th}}{\tau_p}, \quad \left[ \eta_e \frac{1}{qA} - \frac{g_{el} \eta_l}{\tau_{el} qA} \left( \frac{1}{\tau_e} + \frac{1}{\tau_{el}} \right)^{-1} \right] I_{th} \geq \frac{S_{th}}{\tau_p}$$

in which $S_{th}$ is the photon density at the threshold of lasing. Hence, the threshold current for lasing is given by the stringer condition. Mathematically, we obtain

$$I_{th} = \frac{qAS_{th}}{\tau_p} \left( \min \left\{ \eta_h, \eta_e - \frac{g_{el} \eta_l}{\tau_{el}} \left( \frac{1}{\tau_e} + \frac{1}{\tau_{el}} \right)^{-1} \right\} \right)^{-1}$$

Now the photon density at the threshold $S_{th}$ is estimated from the above, yielding the equation

$$\frac{1}{\tau_p} S_{th} = g_{eh}(N_e) \frac{N_h}{\tau_{eh}}$$

And thus the threshold current

$$I_{th} = g_{eh}(N_e) \frac{qAN_h}{\tau_{eh}} \left( \min \left\{ \eta_h, \eta_e - \frac{g_{el} \eta_l}{\tau_{el}} \left( \frac{1}{\tau_e} + \frac{1}{\tau_{el}} \right)^{-1} \right\} \right)^{-1}$$

$$\approx g_{eh}(N_e) \frac{qAN_h}{\eta_h \tau_{eh}}$$